\documentclass[epj]{svjour}
\usepackage{latexsym}
\usepackage{graphics}
\usepackage{cite} 
\usepackage{amsmath}
\usepackage{hyperref}
\usepackage{cite}
\usepackage{amsmath,amssymb}
\usepackage{color}

\usepackage[usenames,dvipsnames]{xcolor}

\begin{document}
\title{Model comparison using supernovae and Hubble parameter data}
\author{Grigoris Panotopoulos
\inst{1} 
\thanks{\emph{E-mail: \ }  \href{mailto:grigorios.panotopoulos@tecnico.ulisboa.pt}{grigorios.panotopoulos@tecnico.ulisboa.pt}}
\and 
\'Angel Rinc\'on
\inst{2}
\thanks{\emph{E-mail: \ } \href{mailto:arrincon@uc.cl}{arrincon@uc.cl}}%
}                     
%
%
\institute{Centro de Astrof{\'i}sica e Gravita{\c c}{\~a}o, Instituto Superior T{\'e}cnico-IST,
Universidade de Lisboa-UL, Av. Rovisco Pais, 1049-001 Lisboa, Portugal.
\and 
Instituto de F\'{i}sica, Pontificia Universidad Cat\'{o}lica de Chile, \mbox{Avenida Vicu\~na Mackenna 4860, Santiago, Chile.}
}
\date{Received: date / Revised version: date}
%
\abstract{
We compare several well-known cosmological dark energy models using observational data coming both from type Ia supernovae and from Hubble parameter measurements. First we use supernovae data to determine the free parameters of each model except for the Hubble constant $H_0$. Then, we use the Hubble parameter measurements to determine $H_0$, and finally the age of the Universe for each model is computed. Our main numerical results are summarized in tables, and they are briefly discussed. The prediction for the decelerating parameter as a function of red-shift for each model is shown in the same figure for comparison. Our numerical results seem to indicate a correlation between the deceleration parameter and the value of the Hubble constant.%
\PACS{
      {PACS-key}{discribing text of that key}   \and
      {PACS-key}{discribing text of that key}
     } 
} 
\maketitle
%

\section{Introduction}

For several decades, from the discovery of the expansion of the Universe by E.~Hubble in 1929 \cite{edwin} until the end of the 90's, it was believed that the Universe was expanding in a decelerating rate driven by radiation and non-relativistic matter. That picture changed in 1998 when two independent groups analysing supernovae (SN) data discovered that the Universe is expanding in an accelerating rate \cite{SN1,SN2}. Nowadays a plethora of well-established observational data coming both from Cosmology and Astrophysics seem to show that the Universe is spatially flat dominated by dark energy (DE) \cite{turner}, which is one of the biggest challenges for modern theoretical cosmology as its origin and nature still remain a mystery. The concordance cosmological model, which is based on cold dark matter and a cosmological constant (or $\Lambda$CDM model), is the most economical one in excellent agreement with current data. Despite its success, however, it suffers from the cosmological constant problem \cite{weinberg}. Therefore, many alternatives have been introduced and investigated over the years in the literature. In general all dark energy models fall into two broad classes, namely either dynamical or geometrical models of dark energy. In the first class one must introduce a new dynamical field to accelerate the Universe, whereas in the second one an alternative theory of gravity is assumed to modify Einstein's General Relativity at cosmological scales. In the first class one finds models such as quintessence \cite{DE1}, phantom \cite{DE2}, quintom \cite{DE3}, tachyonic \cite{DE4} or k-essence \cite{DE5}, while in the second class one finds models such as $f(R)$ theories of gravity \cite{mod1,mod2,HS,starobinsky}, brane-world models \cite{langlois,maartens,dgp} and Scalar-Tensor theories of gravity \cite{BD,leandros,PR}. For an excellent review on the dynamics of dark energy see e.g. \cite{copeland}.

In a recent paper \cite{paperbase} it was pointed out that for the $\Lambda$CDM model, the value of the Hubble constant extracted by the PLANCK Collaboration \cite{planck}, $H_0 = (67-68)~\text{km/(Mpc  sec)}$, was lower than the value obtained by local measurements, $H_0 = (73-74)~\text{km/(Mpc sec)}$ \cite{hubble,recent}. It is known that there is a tension between high red-shift CMB data and low red-shift data, a fact already realized in previous works, see e.g. \cite{tension,tension1,tension2,tension3}. Therefore, the authors of \cite{paperbase} studied a non-standard anisotropic Bianchi-type cosmological model based on the Brans-Dicke theory \cite{BD} in an attempt to alleviate the tension between the PLANCK Collaboration and Hubble constant direct measurements.
 
As there are several dark energy alternatives studied over the years, it would be interesting to see for which models one would obtain similar results and for which models the tension would persist. 
Therefore, in the present work we wish to extend the work of \cite{paperbase} by analysing and comparing several dark energy models using the same data and performing the same two-step analysis as in \cite{paperbase}. As we discuss later on in section 3, an interesting pattern between the Hubble constant and the behaviour of the decelerating parameter is observed. In addition to that, we find that the aforementioned tension is alleviated, in the spirit of \cite{paperbase}, even for $\Lambda$CDM, although in some models it persists.

Our work is organized as follows: after this introduction, we present the dark energy models in section 2, while the numerical results are discussed in the third section. Finally we conclude our work in the last section.

\section{The dark energy models}

We consider a flat Friedmann-Robertson-Walker metric
\begin{equation}
\mathrm{d}s^2 = -\mathrm{d}t^2 + a(t)^2 [\mathrm{d} r^2 + r^2 (\mathrm{d} \theta^2 + \sin^2 \theta \mathrm{d} \varphi^2)]
\end{equation}
with $t$ being the cosmic time and $a(t)$ being the scale factor. In the discussion to follow we shall be needing the Hubble parameter $H \equiv \dot{a}/a$ as a function of red-shift, $z \equiv -1+a_0/a$, where $a_0$ is the present value of the scale factor.
It is convenient to define the normalized density for each fluid component
\begin{equation}
\Omega = \frac{\rho}{\rho_{cr}}
\end{equation}
where $\rho$ is the energy density of the fluid component, and $\rho_{cr}=3H_0/(8 \pi G)$, with $G$ being Newton's constant. Finally, if $H_0=H(z=0)$ is the Hubble parameter, we define the dimensionless function $E(z) \equiv H(z)/H_0$. Then, in the present work we shall consider the following dark energy models:

\begin{itemize}
\item $\Lambda$CDM model: It is the most economical model characterized by a single parameter, namely the matter normalized density $\Omega_m$. The Hubble parameter is given by \cite{nesseris}
\begin{equation}
E(z) = \sqrt{\Omega_m (1+z)^3 + 1-\Omega_m}
\end{equation}
This model may be viewed as a fluid with a constant equation-of-state parameter $w=-1$.

\item $w$CDM model: The equation-of-state $w$ is assumed to be a constant, but not necessarily equal to $-1$. This model slightly extends the $\Lambda$CDM model in a straightforward way, and it is reduced to the previous model when $w = -1$.
It was called "quiessence" in \cite{sahni}, not to be confused with quintessence, which is based on a canonical scalar field. The Hubble parameter is given by \cite{nesseris}
\begin{equation}
E(z) = \sqrt{\Omega_m (1+z)^3 + (1-\Omega_m) (1+z)^{3 (1+w)}}
\end{equation}
with two free parameters, namely the matter normalized density $\Omega_m$ and the dark energy equation-of-state parameter $w$.

\item Linear ansatz: Assuming a dynamical dark energy model characterized by some time varying equation-of-state $w(z)$, one may consider at low red-shift the first order Taylor expansion of $w(z)$ around zero \cite{linear1,linear2,linear3}
\begin{equation}
w(z) = w_0 + w_1 z
\end{equation}
The Hubble parameter is given by \cite{nesseris}
\begin{equation}
E(z) = \sqrt{\Omega_m (1+z)^3 + (1-\Omega_m) f(z)}
\end{equation}
where the function $f(z)$ is given by
\begin{equation}
f(z)  = e^{3 w_1 z} (1+z)^{1+w_0-w_1}
\end{equation}
This model is characterized by 3 free parameters, namely $w_0, w_1, \Omega_m$.

\item Linder ansatz: In this model the dark energy equation-of-state parameter $w(z)$ is considered to be a function of red-shift given by \cite{Linder}
\begin{equation}
w(z) = w_0 + w_1 \frac{z}{1+z}
\end{equation}
This model is reduced to the previous linear model at low red-shift, while at the same time it is well behaved and bounded at high red-shift. It is designed to interpolate between $w_0$ at $z=0$ and $w_0+w_1$ at $z \gg 1$. If it remains in the range $(-1,1)$ it may be described at the Lagrangian level introducing a canonical scalar field with an appropriate self-interaction potential. The Hubble parameter is given by \cite{nesseris}
\begin{equation}
E(z) = \sqrt{\Omega_m (1+z)^3 + (1-\Omega_m) f(z)}
\end{equation}
where the function $f(z)$ is given by
\begin{equation}
f(z) = e^{3 \left(-w_1+\frac{w_1}{1+z}\right)} (1+z)^{1+w_0+w_1}
\end{equation}
This model is characterized by 3 free parameters, namely $w_0, w_1, \Omega_m$.

\item Interacting dark energy: In this model a non-vanishing interaction between dark energy and non-relativistic matter is assumed. The simplest case corresponds to a source term linear in the matter energy density $\rho_m$, $Q=\delta H \rho_m$, with the constant $\delta$ being the strength of the interaction. In this model it is possible to obtain an analytical expression for the Hubble parameter as a function of red-shift, which is computed to be \cite{interacting}
\begin{align}
\begin{split}
E(z)  = & \ \sqrt{ \Omega_X (1+z)^{3 (1+w)} + \frac{1-\Omega_X}{\delta+3w}  
\bigg[\delta (1+z)^{3(1+w)}+3w (1+z)^{3-\delta}\bigg]}
\end{split}
\end{align}
The model is characterized by 3 free parameters, $w, \delta, \Omega_X$, with $w, \Omega_X$ being the dark energy equation-of-state parameter and normalized density, respectively. Clearly, when $\delta=0$ we recover the $w$CDM model.

\item Dvali-Gabadadze-Porrati brane model: In the brane-world idea \cite{langlois,maartens} inspired from Superstring Theory \cite{ST1,ST2}, all known particles and interactions must live on a hyper surface with 3 spatial dimensions, called the brane, while gravity is free to propagate in the bulk with at least one extra transverse dimension. The cosmology of the DGP model \cite{dgp} was studied in \cite{deffayet}, and the Hubble parameter is given by \cite{deffayet}
\begin{equation}
E(z) = \sqrt{\Omega_C} + \sqrt{\Omega_m (1+z)^3 + \Omega_C}
\end{equation}
where $\Omega_C$ is given by
\begin{equation}
\Omega_C = \left( \frac{1-\Omega_m}{2} \right)^2
\end{equation}
and it is characterized by a single parameter $\Omega_m$.

\item The last model considered here is a Scalar-Tensor model that was proposed and studied by us in \cite{PR}, where we showed that the cosmological equations admit an exact power-law solution for the scale factor, $a(t) \sim t^p$. The Hubble parameter is given by
\begin{equation}
E(z) = (1+z)^{1/p}
\end{equation}
and the model is characterized by a single parameter $p$.
\end{itemize}

Finally, the age of the Universe $t_0$ within a certain model can be computed evaluating the integral
\begin{equation}
t_0 = H_0^{-1} \: \int_0^\infty \mathrm{d}z \frac{1}{(1+z) E(z)}
\end{equation}

Although a Lagrangian description of dark energy models would be the ideal one, in practice dark energy parameterizations provide us with a more convenient phenomenological description. It is not uncommon to investigate and constrain the properties of dark energy analysing the predictions of concrete dark energy parameterizations. Just to mention a couple of them, in the past different dark energy parameterizations were studied and compared to SN data \cite{nesseris,grigoris,Magana}. First, in \cite{nesseris} the authors using older SN data (recent back then) compared several dark energy parameterizations, and concluded that models that cross the $w=-1$ line have a better fit to data. The $w$CDM model, the linear ansatz as well as the Linder ansatz, among others, were considered in that work, while the DGP model was considered in \cite{grigoris}. Furthermore, the findings of a more recent work indicate that cosmic acceleration may have slowed down recently \cite{Magana}.

We acknowledge the fact that some of the models considered here are not so popular nowadays. However, we have chosen these particular models due to their simplicity with a twofold goal in mind, namely a) to cover a broad range of different dark energy models (dynamical, interacting, geometrical etc), and b) to see whether or not there is a pattern between the value of the Hubble constant and the decelerating parameter. 

\section{Numerical results}

In this section we perform the numerical analysis, following \cite{paperbase}, in two steps. In particular, first we use the type Ia SN data from the Union 2 compilation \cite{union2} to determine all free parameters, $\Omega_m,w,...$ but the Hubble constant. After that, in the second step, we use Hubble parameter data to determine $H_0$.

The supernovae distance modulus $\mu = m-M$, where $M$ is the absolute and $m$ the apparent magnitude, is given by \cite{copeland,nesseris}
\begin{equation}
\mu(z) = 25 + 5 \log_{10} \left[ \frac{d_L(z)}{\text{Mpc}} \right]
\end{equation}
where the luminosity distance is given by \cite{hogg}
\begin{equation}
d_L(z) = (1+z) \int_0^z \mathrm{d}x \frac{1}{H(x)} 
\end{equation}
The free parameters of each model are determined upon minimization of the $\chi_{SN}^2$ given by
\begin{equation}
\chi_{SN}^2 = X - \frac{Y^2}{Z} + \log_{10} \left( \frac{Z}{2 \pi} \right)
\end{equation}
where $X, Y, Z$ are given by \cite{paperbase}
\begin{eqnarray}
X & = & \sum_i^N \frac{(\mu_{th}(z_i)-\mu_i)^2}{\sigma_{\mu,i}^2} \\
Y & = & \sum_i^N \frac{\mu_{th}(z_i)-\mu_i}{\sigma_{\mu,i}^2} \\
Z & = & \sum_i^N \frac{1}{\sigma_{\mu,i}^2}
\end{eqnarray}
with $N=557$ being the number of the SN data points, and where $\sigma_\mu$ is the error in $\mu$ of the data.

Then we use measurement data of the Hubble parameter shown in the table \ref{hubble} (taken from \cite{paperbase}), and we minimize a new $\chi_H^2$ corresponding to the Hubble parameter
\begin{equation}
\chi_H^2  =  \sum_i^{19} \frac{(H_{th}(z_i)-H_i)^2}{\sigma_{H,i}^2}
\end{equation}
with $\sigma_H$ being the error in $H$ of the data.

We repeat the same steps for all models presented before, and we summarize our main numerical results in the tables~\ref{summary1} and ~\ref{summary2} below. Note that since the number of free parameters differs from one model to another, we have computed $\chi_{\text{min}}^2$ per degree of freedom. The number of degrees of freedom is given simply by $N-n$, where $N$ is the number of data points, and $n$ is the number of the free parameters of the model at hand. The minimum value of $\chi_{\text{min}}^2$ per degree of freedom for each model is shown too. 

We see that as far as the SN data are concerned, $\Lambda$CDM offers us the best fit to the data. The linear ansatz and the Linder ansatz contain the most non-relativistic matter, while DGP and the interacting model contain the least. 
In most of the models discussed here the age of the Universe is comparable to the one corresponding to the $\Lambda$CDM model, $\sim 13$~Gyr, except the ST model and the interacting dark energy model. In the latter case in particular it is found to be too low, given that there are lower limits on the age of the Universe, $t_0 > 12$~Gyr at least \cite{age}. 

Furthermore, we can see that the models considered here may be divided into two classes, namely low Hubble constant, $H_0 \simeq 68~ \text{km}/(\text{Mpc sec})$,
and high Hubble constant, $H_0 \simeq 73~ \text{km}/(\text{Mpc sec})$. In particular, the linear ansatz, the Linder ansatz as well as the interacting dark energy model fall into the first class, where the computed Hubble constant is comparable to the one extracted by the PLANCK Collaboration, while the rest of the models fall into the second class, where the Hubble constant turns out to be comparable to local measurements, or even higher in the ST model. We see that for the models of the second class we obtain results similar to the ones found in \cite{paperbase}, even for $\Lambda$CDM, although in the models of the first class the tension persists.

\begin{figure}
\centering
\resizebox{\textwidth}{!}{%
  \includegraphics{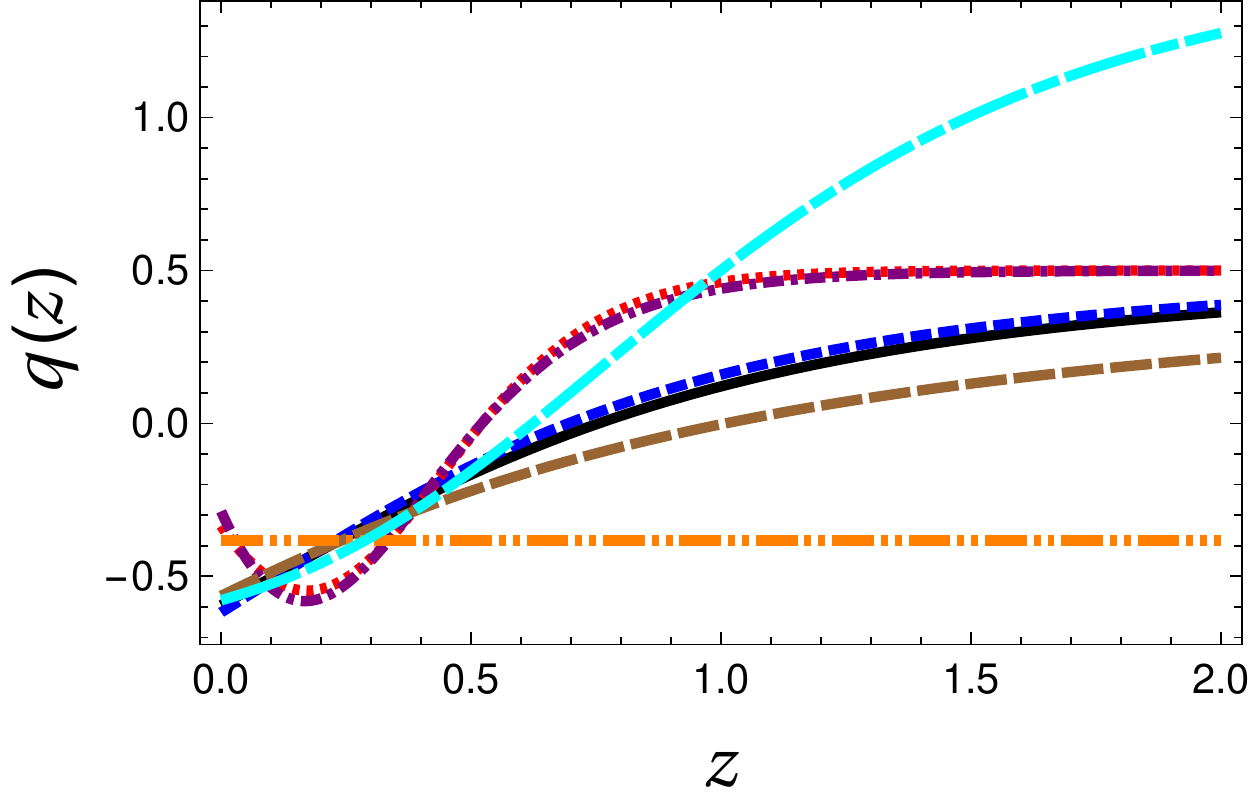}  \ \
  \includegraphics{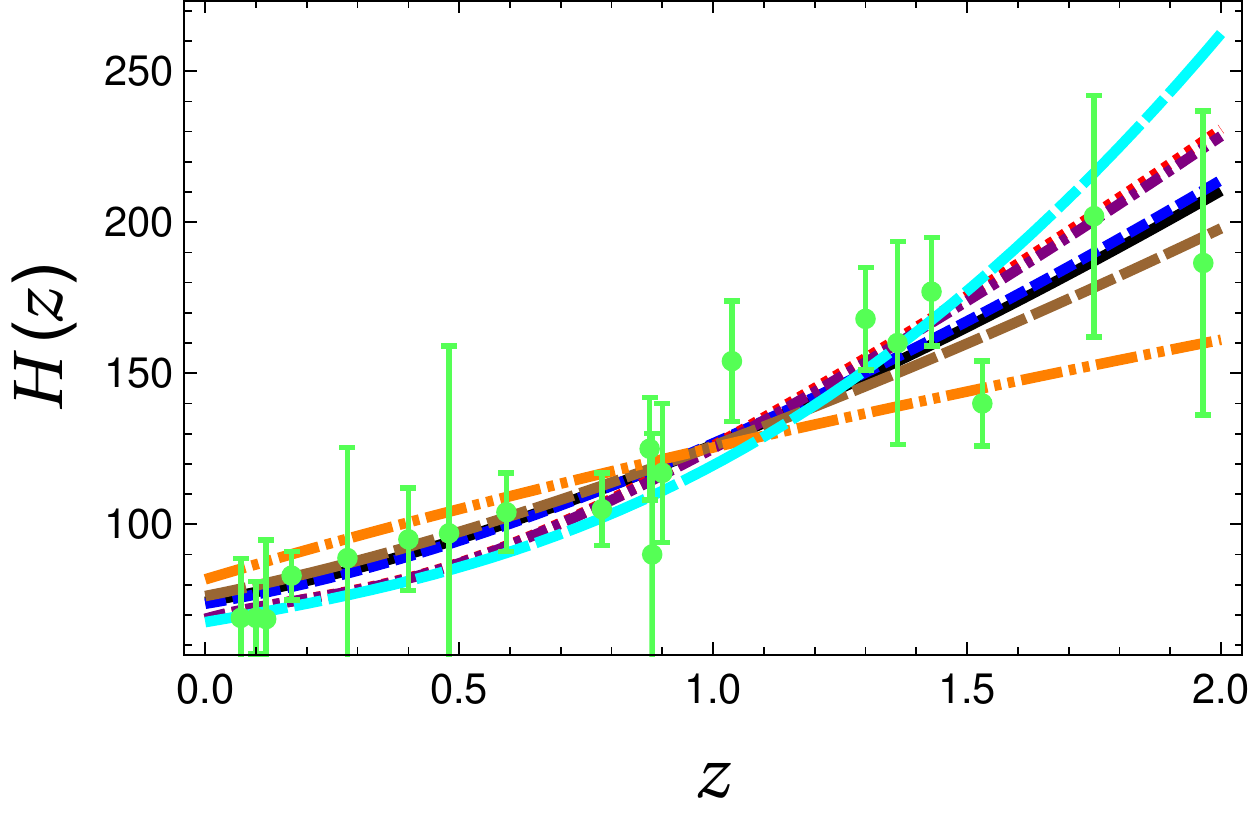}
}
\caption{{\bf Left panel:} 
Deceleration parameter $q$ as a function of the red-shift $z$ for the seven models. {\bf Right panel:} Hubble parameter $H$ as function of the red-shift $z$ for the same models. The data is shown with green dots with the corresponding error bars. The color code is as follows: i) $\Lambda$CDM (solid black line), ii) $w$CDM (dashed blue line),
iii) Linear ansatz (dotted red line), iv) Linder ansatz (dotted-dashed purple line), v) Interacting DE (long-dashed dashed cyan line), vi) DGP (long dashed brown line), and vii) Scalar-Tensor (double-dotted dashed orange).
}
\label{fig:1}       
\end{figure}

Finally, in Fig. \ref{fig:1} (left panel) we show the decelerating parameter $q \equiv -\ddot{a}/(a H^2)$, which is computed to be
\begin{equation}\label{desa}
q(z) = -1 + (1+z) \: \frac{E'(z)}{E(z)}
\end{equation}
for all models presented in the previous section. Furthermore, in the same figure (right panel), we show the Hubble parameter vs red-shift for all models. The data points presented in table \ref{hubble} are shown too. Interestingly enough, by looking at table~\ref{summary2} and at the left panel of fig.~\ref{fig:1} it seems to us that it is possible to correlate the value of the Hubble constant with the deceleration parameter as a function of red-shift. 
In particular, our numerical results seem to indicate that the models with a low Hubble constant $H_0 \simeq 68~ \text{km}/(\text{Mpc sec})$, such as the linear ansatz, the Linder ansatz as well as the interacting DE model, exhibit a cosmic slowing down of acceleration very recently.

\begin{table}
\centering
  \caption{Hubble parameter measurements}
  \begin{tabular}{ccc}
  \hline
$z$ & $H(z)$ & $\sigma_H$ \\
\hline
0.07 &  69  &  19.6 \\
0.1	 &  69  &  12   \\
0.12	 & 68.6 &  26.2 \\
0.17	 &  83  &  8    \\
0.28 &  88.8  &  36.6  \\
0.4  &  95  &  17  \\
0.48 &  97  &  62  \\
0.593 & 104  &  13  \\
0.781 & 105  & 12  \\
0.875 &  125  & 17  \\
0.88 &  90  &  40 \\
0.9  &  117  &  23 \\
1.037 &  154  &  20  \\
1.3  &  168  &  17 \\
1.363 &  160  &  33.6 \\
1.43  &  177  &  18 \\
1.53  &  140  &  14 \\
1.75  &  202  &  40  \\
1.965 &  186.5  &  50.4  \\
\hline
\end{tabular}
\label{hubble}
\end{table}

\begin{table*}
\centering
  \caption{Summary of our main numerical results I}
  \begin{tabular}{ccccc}
  \hline
Model & Parameter 1 & Parameter 2 & Parameter 3 & $\chi_{\text{min,SN}}^2/(\text{d.o.f})$  \\
\hline
$\Lambda$CDM & $\Omega_m=0.27$ & - & -  & 0.979 \\
$w$CDM & $\Omega_m=0.29$  & $w=-1.05$ & - & 0.981  \\
Linear ansatz &  $\Omega_m=0.42$ & $w_0=-0.96$ & $w_1=-3.9$ & 0.981  \\
Linder ansatz &  $\Omega_m=0.41$ & $w_0=-0.89$ & $w_1=-5.11$ & 0.981  \\
Interacting DE & $\Omega_X=0.91$ & $\delta=-2.12$ & $w=-0.79$ & 0.982 \\
DGP & $\Omega_m=0.17$ & - & - & 0.980 \\
Scalar-Tensor & $p=1.62$ & - & - & 0.991 \\
\hline
\end{tabular}
\label{summary1}
\end{table*}

\begin{table*}
\centering
  \caption{Summary of our main numerical results II}
  \begin{tabular}{cccc}
  \hline
Model & $H_0~\text{km}/(\text{Mpc  sec)}$ & $\chi_{\text{min},H}^2/(\text{d.o.f})$ & $t_0$~[Gyr] \\
\hline
$\Lambda$CDM   & 74.30 &  0.565  & 13.05  \\
$w$CDM         & 73.68 &  0.573  & 13.03  \\
Linear ansatz  & 68.55 &  0.782  & 13.10  \\
Linder ansatz  & 68.68 &  0.768  & 13.09  \\
Interacting DE & 67.60 &  1.098  & 11.91  \\
DGP            & 76.22 &  0.564  & 13.34  \\
Scalar-Tensor  & 81.76 &  0.973  & 19.29  \\
\hline
\end{tabular}
\label{summary2}
\end{table*}


\section{Conclusions}

To summarize, in this article we have compared several well-known cosmological dark energy models using observational data coming both from type Ia supernovae and from local measurements of the Hubble parameter. Following a previous work we have determined the free parameters of the models as well as the Hubble constant in a two-step procedure. First we used SN data to determine the free parameters of each model except for the Hubble parameter $H_0$. Then, we used the Hubble parameter measurements to determine $H_0$, and finally the age of the Universe for each model was computed. We have shown the main numerical results in tables, and they are briefly discussed. Finally, the prediction for the decelerating parameter as a function of red-shift for each model is shown in the same figure for comparison. An interesting pattern between the value of the Hubble constant and the deceleration parameter is observed.


\section*{Acknowledgments}

\noindent
We are grateful to L. Perivolaropoulos for emailing us the Union 2 SN data.
The work of A.R. was supported by the CONICYT-PCHA/Doctorado Nacional/2015-21151658. 
G.P. thanks the Funda\c c\~ao para a Ci\^encia e Tecnologia (FCT), Portugal, for the financial support to the Center for Astrophysics and Gravitation-CENTRA, Instituto Superior T\'ecnico, Universidade de Lisboa, through the Grant No. UID/FIS/00099/2013.

%
%
%

%
%

\end{document}